\documentclass[%
 reprint,
 amsmath,amssymb,
 aps,
]{revtex4-1}

\usepackage{graphicx}
\usepackage{dcolumn}
\usepackage{bm}
\usepackage{xcolor}

\newcommand{\oex}{\omega_{\rm ex}}
\newcommand{\dfracp}[2]{\dfrac{\partial #1}{\partial #2}}
\newcommand{\bh}{\boldsymbol{h}}
\newcommand{\norm}[1]{\left\lVert #1\right\Vert}
\newcommand{\wt}{\widetilde}
\newcommand{\wh}{\widehat}

\begin{document}


\title{Reconstruction of  Phase Dynamics from Macroscopic Observations Based on Linear and Nonlinear Response Theories}

\author{Yoshiyuki Y. Yamaguchi$^{1}$}
\email{yyama@amp.i.kyoto-u.ac.jp}
\author{Yu Terada$^{2,3,4}$}
\email{yuterada@ucsd.edu}
\affiliation{
  $^{1}$Department of Applied Mathematics and Physics, Graduate School of Informatics, Kyoto University, Kyoto 606-8501, Japan\\
  $^{2}$Neurobiology Section, Division of Biological Sciences, University of California San Diego, La Jolla, CA 92093, United States of America\\
$^{3}$Institute for Physics of Intelligence, Department of Physics Graduate School of Science, The University of Tokyo 7-3-1 Hongo, Bunkyo-ku, Tokyo 113-0033, Japan\\
$^{4}$Laboratory for Neural Computation and Adaptation, RIKEN Center for Brain Science, 2-1 Hirosawa, Wako, Saitama 351-0198, Japan} 


\begin{abstract}
  We propose a novel method to reconstruct phase dynamics equations
  from responses in macroscopic variables to weak inputs.
  Developing linear and nonlinear response theories in coupled phase-oscillators,
  we derive formulae which connect the responses with the system parameters
  including the time delay in interactions.
  We examine our method by applying it to two phase models,
  one of which describes a mean-field network
  of the Hodgkin--Huxley type neurons with a nonzero time delay. 
  The method does not require much invasiveness nor microscopic observations,
  and these advantages highlight its broad applicability in various fields.
\end{abstract}


\maketitle


Rhythmical phenomena have been ubiquitously observed in nature as well as in engineering systems and attracted a wide spectrum of interests
\cite{winfree-01,strogatz-03,pikovsky-01}.
Specific rhythmical dynamics are believed to play crucial functional roles in information processing of the brain \cite{palmigiano-17,buzsaki-13}.
Theoretical analysis have contributed to understanding the nature of interacting rhythmical systems.
One significant success in theoretical researches is the phase reduction, which reduces a high-dimensional rhythmic dynamical system to a one-dimensional phase-oscillator system
by eliminating the other nonessential degrees of freedom
\cite{kuramoto-03,nakao-16,kuramoto-19}.
In this framework, a collective system of interacting units is described by a coupled phase-oscillator system,
which consists of the natural frequency distribution, coupling function, and time delay in interactions.
A dynamical system behind an observed rhythmic phenomenon in the real world
is mostly, however, unknown,
while the knowledge helps to profoundly understand, predict, and control it.
This means high demand to specify the underlying coupled phase-oscillator system.

As the reconstruction is a central issue in coupled phase-oscillator systems,
many works have proposed reconstruction methods
\cite{galan-ermentrout-urban-05,miyazaki-kinoshita-05,tokuda-07,kralemann-07,kralemann-08,penny-09,stankovski-12,ota-14,pikovsky-18,mori-22}.
However, there are mainly two rooms that should be addressed.
The first is the assumption of accessibility to individual elements. 
The previous works assume that time series of almost all elements are available, which implausible in some situations.
For example, with electroencephalogram or functional magnetic response imaging signals, we can obtain only mesoscopic or macroscopic activity of the nervous systems.
The second is the inference of the time delay.
The existence of the time delay is in principle inevitable in real systems,
and can drastically change dynamics
\cite{yeung-strogatz-99,montbrio-pazo-schmidt-06}.
It is therefore a next step to develop a method that can be implemented with unknown interaction delay.

Here, we utilize the linear response theory
for coupled phase-oscillator systems
\cite{sakaguchi-88,daido-15,terada-yamaguchi-19}
with the aid of a nonlinear response theory.
We apply weak external forces into a system,
and observe asymptotic responses of order parameters, which are macroscopic variables.
We note that it does not require time series of individual elements
and that the time delay is tractable.
Further, applied external forces are assumed substantially weak, since we focus on a regime where the linear response theory is valid.
This assumption brings another advantage that our approach possesses,
because strong inputs into a system may cause an undesirable change in states
of a system.
The essential assumptions on models are
that the system has the mean-field, all-to-all homogeneous interactions
and that the system lies in the nonsynchronized state.
For the first assumption, it is worth remarking that
the all-to-all interaction may not be extremely special,
because the criticality in the small-world network
\cite{watts-strogatz-98}
belongs to the universality class of the all-to-all interaction
\cite{hong-choi-kim-02,yoneda-harada-yamaguchi-20}.
The mean-field analysis employed here could be extended by assuming statistics in couplings \cite{daido-87,ichinomiya-03}.
The second assumption comes from the effectiveness of linear response theory developed in \cite{terada-yamaguchi-19} and here.

Based on the phase reduction \cite{hoppensteadt-97}
and following the first assumption,
we describe the underlying coupled phase-oscillator system by
\begin{align}
  \frac{d\theta_{j}}{dt}
  = \omega_{j}
  +\dfrac{1}{N} \sum_{k=1}^{N}\Gamma\left(\theta_{j}(t)-\theta_{k}(t-\tau)\right)
  +H(\theta_{j}(t),t;\oex).
  \label{eq:model}
\end{align}
The variable $\theta_{j}(t)$ represents the phase of the $j$th oscillator at time $t$,
the constant $\omega_j$ is the natural frequency following the natural frequency distribution $g(\omega)$,
the function $\Gamma$ represents the coupling function,
the constant $\tau$ is the time delay for the coupling.
The function $H$ represents the external force
and the constant $\oex$ is its frequency.
The system parameters $g(\omega)$, $\Gamma$, and $\tau$
are intrinsically determined but unknown,
and we will infer them from observation of responses to the external force $H$
by varying the controllable frequency $\oex$.
The coupling function $\Gamma(\theta)$ is $2\pi$-periodic
and is expanded into the Fourier series as
\begin{align}
  \Gamma\left(\theta\right)= -\sum_{m=1}^{\infty}K_m\sin\left(m\theta+\alpha_m\right),
  \label{eq:coupling_functions}
\end{align}
where $K_m$ is the coupling strength and $\alpha_m$ is the phase-lag parameter
for the $m$th Fourier component of $\Gamma(\theta)$.
We here apply the external force as
\begin{equation}
  H\left(\theta, t; \oex \right)
  = - \Theta(t) \sum_{m=1}^{\infty} h_{m}\sin\left[m\left(\theta-\oex t\right)\right],
  \label{eq:external_force}
\end{equation}
where $h_{m}$ is the amplitude of the $m$th mode.
The function $\Theta(t)$ is the unit step function:
The external force is off for $t<0$ and kicks in at $t=0$.

The dynamics \eqref{eq:model} are described in the limit $N\to\infty$
by the equation of continuity \cite{lancellotti-05}
governing $F(\theta,\omega,t)$, which
is the probability density function at the time $t$
and normalized as $\int_{-\infty}^{\infty}d\omega\int_{0}^{2\pi} d\theta~ F(\theta,\omega,t)=1$.
The nonsynchronized state specified as $F_{0}(\omega)=g(\omega)/(2\pi)$, which corresponds to the uniform distribution over $\theta$,
is a stationary solution to the equation of continuity. 
The order parameters, whose responses we observe, are defined by \cite{daido-92}
\begin{equation}
  z_{n}(t)
  = \int_{-\infty}^{\infty} d\omega \int_{0}^{2\pi} d\theta~ e^{in\theta} F(\theta,\omega,t).
\end{equation}

Assuming that the external force $\bh=(h_{1},h_{2},\cdots)$ is sufficiently small,
we perturbatively analyze the equation of continuity
by using the Fourier transform in $\theta$ and the Laplace transform in $t$.
Supposing that $F_{0}$ is stable,
we obtain the asymptotic evolution of $z_{n}(t)$ in the linear regime as
$e^{-in\oex t} z_{n}(t) \xrightarrow{t\to\infty}
\chi_{n}(\oex) h_{n} + O(\norm{\bh}^{2})$,
where we suppose $n>0$ hereafter \cite{terada-yamaguchi-19}.
Smallness of $\bh$ ensures that observation of $e^{-in\oex t}z_{n}$
provides a good approximation of $\chi_{n}(\oex)h_{n}$.
Moreover, if we apply $h_{m}~(m>0)$ and observe $e^{-in\oex t}z_{n}~(n\neq m)$,
then we have a nonlinear response of order $O(\norm{\bh}^{2})$.
Our goal is to obtain formulae that allow to reconstruct $\tau$, $K_{m}$'s, $\alpha_{m}$'s, and $g(\omega)$
from observation date of $\{\chi_{n}(\oex)\}$ and nonlinear responses
for a set of external frequency,
$\oex\in\{\oex^{1},\cdots,\oex^{S}\}$, where $\oex^{1}<\cdots<\oex^{S}$.
We call a sampling reliable,
if the range $\oex^{S}-\oex^{1}$ is sufficiently large
and the gaps $\oex^{i+1}-\oex^{i}$ are sufficiently small.

The susceptibility $\chi_{n}(\oex)$ of the linear response reads \cite{supplement}
\begin{equation}
  \label{eq:chin}
  \chi_{n}(\oex)
  = \dfrac{\mathcal{G}(\oex)}{2-L_{n}(\oex)\mathcal{G}(\oex)}
  \quad (n>0),
\end{equation}
where $L_{n}(\oex) = K_{n} e^{-i(\alpha_{n}+n\oex\tau)}$ and
$\mathcal{G}(\oex)=\pi g(\oex) + i~{\rm PV}
\int_{-\infty}^{\infty} d\omega~ g(\omega)/(\omega-\oex)$.
The symbol PV indicates the Cauchy principal value.
We remark that $\mathcal{G}(\oex)$ does not depend on the mode number $n$.
Thanks to this independence,
once we obtain one of $L_{m}$'s, say $L_{n}$,
the other coefficients are obtained thought the relation
\begin{equation}
  \label{eq:LmLn}
  L_{m}(\oex) - L_{n}(\oex) = \dfrac{1}{\chi_{n}(\oex)} - \dfrac{1}{\chi_{m}(\oex)}.
\end{equation}
This is the key relation in our method.
An obtained $L_{m}$ infers the natural frequency distribution $g(\omega)$
from observation of the susceptibility $\chi_{m}(\oex)$ as
\begin{equation}
  \label{eq:gomega-estimation}
  g(\omega)
  = \dfrac{1}{\pi} {\rm Re}~\mathcal{G}(\omega)
  = \dfrac{1}{\pi} {\rm Re} \left[ \dfrac{2\chi_{m}(\omega)}{1+L_{m}(\omega)\chi_{m}(\omega)} \right].
\end{equation}
Our method is twofold: inference of $\tau$ ({\bf Procedure-1}) and the others
({\bf Procedure-2}).
The latter is further decomposed into the two cases of
$\tau>0$ ({\bf Procedure-2A}) and $\tau=0$ ({\bf Procedure-2B}).

{\bf Procedure-1} performs a finite Fourier transform
\begin{equation}
  \label{eq:Lmn}
  \begin{split}
    L_{mn}(t)
    & = \dfrac{1}{\oex^{S}-\oex^{1}}
    \int_{\oex^{1}}^{\oex^{S}} [L_{m}(\oex)-L_{n}(\oex)] e^{i\oex t} d\oex. \\
  \end{split}
\end{equation}
If the sampling of $\oex$ is perfectly reliable
so as to reproduce the integral of \eqref{eq:Lmn}
in the limit $\oex^{S}-\oex^{1}\to\infty$, we have
$L_{mn}(t)
  \xrightarrow{\oex^{S}-\oex^{1}\to\infty}
  K_{m}e^{-i\alpha_{m}} \delta_{t,m\tau}
  - K_{n}e^{-i\alpha_{n}} \delta_{t,n\tau}$,
where $\delta_{t,t'}$ is the Kronecker delta.
The absolute value $|L_{mn}(t)|$ has one ($\tau=0$) or two ($\tau\neq 0)$ peaks
at $t=m\tau$ and $t=n\tau$,
and the peak positions infer the time delay $\tau$.
An actual sampling induces two types of errors from the above limit: 
One comes from boundedness of $\oex^{S}-\oex^{1}$,
and the other from finiteness of the sample number.
The latter type concerns errors of the numerical integration.
Nevertheless, large peaks appear at $t=m\tau$ and $t=n\tau$
if the sampling is sufficiently reliable, and $K_{m}$ and $K_{n}$
are sufficiently large comparing with the errors.

{\bf Procedure-2A} uses the relation $L_{mn}(m\tau)=K_{m}e^{-i\alpha_{m}}$
under a reliable sampling of $\oex$
to infer $K_{m}$ and $\alpha_{nm}$.
They with $\tau$ give the factor $L_{m}(\omega)$,
and the natural frequency distribution $g(\omega)$ is inferred
by \eqref{eq:gomega-estimation}.
We remark that we solely used linear responses up to this procedure.

{\bf Procedure-2B} is for $\tau=0$, since the peak at $t=0$ mixes
the modes $m$ and $n$, $L_{mn}(0)=K_{m}e^{-i\alpha_{m}}-K_{n}e^{-i\alpha_{n}}$.
The linear equations for $K_{m}e^{-i\alpha_{m}}~(m=1,2,3)$
obtained from $L_{12}(0),L_{13}(0)$, and $L_{23}(0)$, for instance, are degenerate.
We thus use a nonlinear response to infer, for example, $L_{1}$:
$z_{2}$ in $O(\norm{\bh}^{2})$ can be observed
by applying the external force in the first mode $\bh=(h_{1},0,0,\cdots)$
as $e^{-i2\oex t}z_{2}(t)\xrightarrow{t\to\infty} \chi_{2}^{11}(\oex)h_{1}^{2}$.
The nonlinear response coefficient is theoretically obtained as \cite{supplement}
\begin{equation}
  \label{Eq:chi211}
  \chi_{2}^{11}(\oex)
  = \dfrac{2i\mathcal{G}'(\oex)}
  { [2-L_{2}\mathcal{G}(\oex)][2-L_{1}\mathcal{G}(\oex)]^{2} },
\end{equation}
where $\mathcal{G}'(\oex)$ is the derivative of $\mathcal{G}(\oex)$
with respect to $\oex$.
Solving \eqref{Eq:chi211} we have one expression of $\mathcal{G}'(\oex)$.
We independently have another expression of $\mathcal{G}'(\oex)$
through solving \eqref{eq:chin} by $\mathcal{G}$ and derivating it.
The combination of the above two expressions of $\mathcal{G}'(\oex)$ gives
\begin{equation}
  \label{eq:L1}
  L_{1} = K_{1} e^{-i\alpha_{1}}
  = \dfrac{2\chi_{2}^{11}(\oex)}{i\chi_{2}(\oex)\chi_{1}'(\oex)}
  - \dfrac{1}{\chi_{1}(\oex)}
\end{equation}
for $\tau=0$ \cite{supplement}.
We take the average over $S$ estimated values of $L_{1}$
from $\oex^{1},\cdots,\oex^{S}$.
The other coefficients $L_{m}~(m>1)$ are estimated from \eqref{eq:LmLn}
by taking the average.
We remark that {\bf Procedure-2B} is also applicable for $\tau>0$,
where $L_{1}$ is obtained as a solution to a quadratic equation.
However, {\bf Procedure-2A} provides higher performance in inference for a nonzero
time-delay case as compared in an application \cite{supplement}.

By employing the theory developed above, we tackle a reconstruction problem
in two models:
{\bf Model-1} has a delay, that is, $\tau>0$ and {\bf Procedure-2A} is applied,
while {\bf Model-2} does not and {\bf Procedure-2B} is in use.
Their system parameters are arranged in Table \ref{tab:MainResult}.
Numerical simulations of \eqref{eq:model} are performed
in the use of the second-order Runge-Kutta algorithm
with the time step $\Delta t=0.01$.
Responses of order parameters are obtained as the average in the time interval $(50,150]$.
The number of oscillators is $N=10^{5}$.
All the numerical simulations are performed by activating
only one mode in $\bh$ with strength $0.1$:
$h_{m}=0.1$ and $h_{n}=0~(n\neq m)$ for the $m$th mode.
This strength is sufficiently small for the linear response
but sufficiently large for overcoming finite-size fluctuation
of order $O(1/\sqrt{N})$ by the second-order response of order $O(\norm{\bh}^{2})$.

\begin{table}
  \centering
  \caption{True and inferred parameter values of {\bf Model-1} and {\bf Model-2}.
    The inferred values are given for each sample set.
    NI means noninferred values,
    because there is no clear peak around $t=3\tau$ in neither $|L_{34}|$ nor $|L_{35}|$.
    {\bf Procedure-1} implies that $K_{4}$ should be sufficiently small
    from absence of clear peak of $|L_{45}(t)|$ [see Fig.~\ref{fig:Example2-Lmn}(d)].
  }
  \begin{tabular}{l|l|llllll}
    \hline
    {\bf Model-1} & $\tau$ & $K_{1}$ & $\alpha_{1}$ & $K_{2}$ & $\alpha_{2}$ & $K_{3}$ & $\alpha_{3}$ \\ 
    \hline
    Truth & 2 & 1.379 & 0.7884 & 0.568 & -3.0316 & 0.154 & -0.7546 \\
    $\Omega_{1}^{50}$ & 1.987 & 1.383 & 0.820 & 0.596 & -3.016 & 0.153 & -0.864 \\
    $\Omega_{1}^{25}$ & 1.995 & 1.381 & 0.793 & 0.582 & -3.111 & NI & NI \\
    \hline
    \hline
    {\bf Model-2} & $\tau$ & $K_{1}$ & $\alpha_{1}$ & $K_{2}$ & $\alpha_{2}$ \\
    \hline
    Truth & 0 & 1 & 1 & 0 & 0 \\
    $\Omega_{2}^{81}$ & 0.001 & 0.958 & 1.001 & 0.044 & -2.119 \\
    $\Omega_{2}^{41}$ & -0.001 & 1.063 & 0.497 & 0.521 & -0.706 \\
    \hline
  \end{tabular}
  \label{tab:MainResult}
\end{table}

\begin{figure}[h]
  \centering
    \includegraphics[width=9cm]{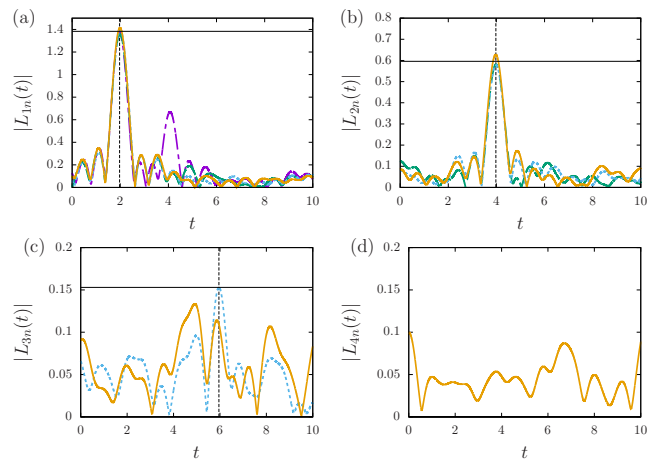}
    \caption{{\bf Procedure-1} in {\bf Model-1}.
      $|L_{mn}(t)|$ \eqref{eq:Lmn}  computed from the sample set $\Omega_{1}^{50}$.
      (a) $m=1$ and $n\in\{2,3,4,5\}$.
      (b) $m=2$ and $n\in\{3,4,5\}$.
      (c) $m=3$ and $n\in\{4,5\}$.
      (c) $m=4$ and $n\in\{5\}$.
      The lines are
      $n=2$ (purple chain),
      $n=3$ (green broken),
      $n=4$ (blue dotted),
      and $n=5$ (orange solid).
      The vertical dashed black lines mark the inferred time-delay $m\tau$,
      and the horizontal solid black lines the inferred $K_{m}$.}
  \label{fig:Example2-Lmn}
\end{figure}

\begin{figure}[h]
  \centering
  \includegraphics[width=9cm]{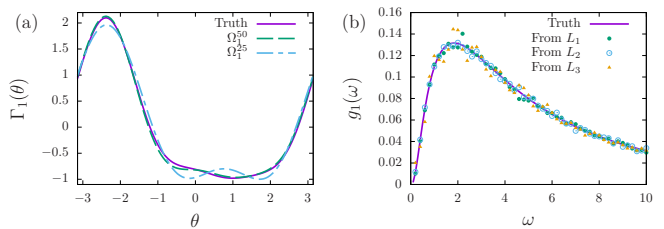}
  \caption{
    Comparison between the truth (purple solid line)
    and the inference in {\bf Model-1} having $\tau>0$.
    (a) The coupling function $\Gamma_{1}(\theta)$.
    The sample sets are $\Omega_{1}^{50}$ (green broken line)
    and $\Omega_{1}^{25}$ (blue chain line).
    (b) The natural frequency distribution $g_{1}(\omega)$ \eqref{eq:case2-gomega}
    obtained from the inferred $L_{1}$ (green filled circles), $L_{2}$ (blue open circles),
    and $L_{3}$ (orange triangles) by \eqref{eq:gomega-estimation}.
    The sample set is $\Omega_{1}^{50}$.
  }
  \label{fig:Example2}
\end{figure}

{\bf Model-1} is motivated by neurobiological systems
and is connected directly to a network of the Hodgkin--Huxley neurons.
As in \cite{hansel-93,hansel-95}, the Fourier components of the modes $m~(m\geq 4)$
are zero.
The time delay is set as $\tau=2$, which is compatible with experimental observations \cite{izhikevich-06}.
Taking another experimental observation \cite{buzsaki-14} into account,
we assume the log-normal natural frequency distribution
\begin{equation}
  \label{eq:case2-gomega}
  g_{1}(\omega) = \dfrac{1}{\omega\sqrt{2\pi\sigma_{1}^{2}}}
  \exp\left[ - \dfrac{(\ln\omega-\mu_{1})^{2}}{2\sigma_{1}^{2}} \right]
\end{equation}
with $\mu_{1}=\ln 5$ and $\sigma_{1}=1$.
The external frequency is sampled from the interval $[0.2, 10]$
with the step $\Delta\oex=0.2$ for the sample set $\Omega_{1}^{50}$ ($S=50$),
and $\Delta\oex=0.4$ for the set $\Omega_{1}^{25}$ ($S=25$).
We start from {\bf Procedure-1}.
We approximately compute $L_{mn}(t)$ \eqref{eq:Lmn} by using the midpoint algorithm,
where a sampling point $\oex^{i}$ is the midpoint.
Absolute values $|L_{mn}(t)|$ for the set $\Omega_{1}^{50}$ are reported 
in Fig.~\ref{fig:Example2-Lmn}.
We obtain the estimate $\tau=1.987$
by taking the average over the largest peak positions
for the pairs $(m,n)=(3,4)$ and $(m',n')~(m'=1,2; n'=m'+1,\cdots,5)$.
A graph should have two large peaks at $t=m\tau$ and $t=n\tau$,
but some peaks are not visible in Fig.~\ref{fig:Example2-Lmn}.
No clear peak at $t=n\tau$ implies that $K_{n}$ is smaller than the error level.
Indeed, no clear peak of $|L_{45}(t)|$ in Fig.~\ref{fig:Example2-Lmn}(d)
is consistent with $K_{4}=K_{5}=0$.
{\bf Procedure-2A} infers the coefficients $L_{m}$'s
from the value of $L_{mn}(t)$ at the peak position,
where the above mentioned pairs are in use to take the average.
Performing the same procedure but using the set $\Omega_{1}^{25}$,
we obtain another set of inferences.
The inferences are compared with the true values in Table \ref{tab:MainResult}.
The coupling function $\Gamma_{1}(\theta)$ is directly obtained from $L_{m}$'s,
and the natural frequency distribution $g_{1}(\omega)$
is inferred through the relation \eqref{eq:gomega-estimation}.
They are in good agreement with the true ones for the set $\Omega_{1}^{50}$
as exhibited in Fig.~\ref{fig:Example2}.
Increasing the number of samples improves the inference,
because the sampling set becomes more reliable.

{\bf Model-2} is the Sakaguchi--Kuramoto model \cite{sakaguchi-86} which is specified by the parameter set
$(K_{1},\alpha_{1})=(1,1)$ and the other Fourier modes are zero.
To demonstrate the ability of the proposed method for general natural frequency distributions,
a nonunimodal and asymmetric natural frequency distribution is assumed as
\begin{equation}
  \label{eq:case1-gomega}
  g_{2}(\omega)
  = \dfrac{ae^{-(x-\mu_{2})^{2}/(2\sigma_{2}^{2})} + (1-a) e^{-(x+\mu_{2})^{2}/(2\sigma_{2}^{2})}}{\sqrt{2\pi}},
\end{equation}
where $a=0.8,~ \mu_{2}=2$, and $\sigma_{2}=1$.
The external frequency is sampled from $[-4,4]$
with the step $\Delta\oex=0.1$ for the sample set $\Omega_{2}^{81}$ ($S=81$)
and $\Delta\oex=0.2$ for the set $\Omega_{2}^{41}$ ($S=41$).
To compute the derivative $\chi_{1}'(\oex)$, we use the central difference
except for the head and the end points, namely $\oex^{1}$ and $\oex^{S}$,
for which the forward and backward differences are in use, respectively.

\begin{figure}[t]
  \centering
    \includegraphics[width=8.7cm]{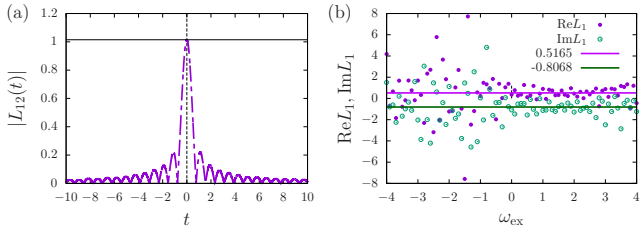}
    \caption{{\bf Model-2}.
      (a) {\bf Procedure-1}.
      The peak position is $\tau=0.001$ and the peak height is $1.014$.
      (b) {\bf Procedure-2B} to infer $L_{1}$
      by \eqref{eq:L1} for each external frequency $\oex$.
      The real part ${\rm Re}L_{m}$ (purple filled circles)
      and the imaginary part ${\rm Im}L_{m}$ (green open circles).
      The purple and green horizontal solid lines mark the averaged values.
      The sample set is $\Omega_{2}^{81}$.}
  \label{fig:Example1-L1L2}
\end{figure}

\begin{figure}[t]
  \centering
  \includegraphics[width=9cm]{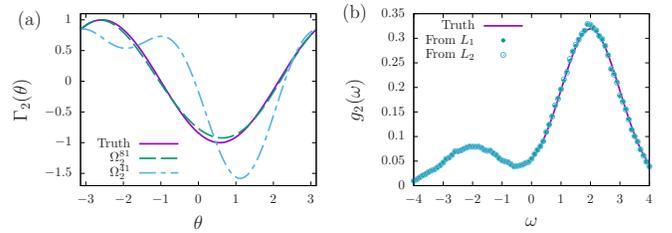}
  \caption{Comparison between the truth (purple solid line) and the inference
    in {\bf Model-2} having $\tau=0$.
    (a) The coupling function $\Gamma_{2}(\theta)$.
    The sample sets are $\Omega_{2}^{81}$ (green broken line)
    and $\Omega_{2}^{41}$ (blue chain line).
    (b) The natural frequency distribution $g_{2}(\omega)$ \eqref{eq:case1-gomega}
    obtained from the inferred $L_{1}$ (green filled circles) and $L_{2}$ (blue open circles)
    through \eqref{eq:gomega-estimation}. The sample set is $\Omega_{2}^{81}$.
  }
  \label{fig:Example1}
\end{figure}

From now on, we concentrate on inferences of $L_{1}$ and $L_{2}$.
{\bf Procedure-1} confirms that $|L_{12}(t)|$ has a large peak at $t=0.001$
[see Fig.~\ref{fig:Example1-L1L2}(a)],
and hence we conclude no time-delay, $\tau=0$.
The peak height $1.014$ corresponds to $|K_{1}e^{-i\alpha_{1}}-K_{2}e^{-i\alpha_{2}}|$,
and the fact $K_{2}=0$ implies that the peak height approximately
infers the value of $K_{1}=1$.
However, we do not know the value of $K_{2}$ a priori,
and we cannot determine $K_{1}$ yet.
We thus use {\bf Procedure-2B}, \eqref{eq:L1}, for inferring $L_{1}$,
and \eqref{eq:LmLn} for $L_{2}$.
They are obtained as functions of $\oex$,
and $L_{1}(\oex)$ is reported in Fig.~\ref{fig:Example1-L1L2}(b).
We determine the inferred values of the constants $L_{1}$ and $L_{2}$
by taking the average over $\oex$,
and the constants $K_{m}$ and $\alpha_{m}~(m=1,2)$ from the averaged $L_{m}$.
The inferred values are arranged in Table \ref{tab:MainResult}.
The set $\Omega_{2}^{81}$ infers good values,
while the set $\Omega_{2}^{41}$ does not provide good inferences,
due to the lack of precision in computation of the derivative $\chi_{1}'(\oex)$.
The inferred coupling function $\Gamma_{2}$
and the natural frequency distribution $g_{2}(\omega)$
agree with the true ones as reported in Fig.~\ref{fig:Example1}.

In summary, we proposed a method to reconstruct 
the underlying coupled phase-oscillator model of a collective rhythmic system
by observing responses in order parameters
to a weak external force with varying its frequency.
Non-invasivity is respected due to weakness of the external force,
and we do not need to know activity of individual elements of the system.
The proposed method is examined through numerical simulations in two models.
The unknown system parameters including the time delay in interactions
have been successfully inferred,
when the sampling of the external frequency lies on a sufficiently large range
with sufficiently small gaps.
Finally, we remark on potential directions of development:
extensions to synchronized states, to noisy systems, and to network systems.

Y.Y.Y. acknowledges the support of JSPS KAKENHI
Grants No. 16K05472 and No. 21K03402.
Y.T. is supported by the Special Postdoctoral Research Program at RIKEN
and JSPS KAKENHI Grant No. 19K20365.

\newpage
\appendix
\begin{widetext}

\section{Linear and nonlinear response theories}

\subsection{Equations to analyze}

We consider the equation of motion
\begin{align}
  \frac{d\theta_{j}}{dt}
  = \omega_{j}
  +\dfrac{1}{N} \sum_{k=1}^{N}\Gamma\left(\theta_{j}(t)-\theta_{k}(t-\tau)\right)
  +H(\theta_{j},t;\oex),
  \qquad
  (j=1,\cdots,N).
  \label{eq:model}
\end{align}
The variable $\theta_{j}$ is the phase of the $j$th phase-oscillator.
The natural frequency $\omega_{j}$ follows the natural frequency distribution $g(\omega)$.
The function $\Gamma$ is the coupling function and
the constant $\tau$ is the time delay.
We assume that the external force $H$ is sufficiently small, i.e.
$\norm{H}\ll 1$, where $\norm{H}$ is a certain norm of the function $H$.
Dynamics of \eqref{eq:model} are described in the limit $N\to\infty$
by the equation of continuity
\begin{equation}
  \dfracp{F}{t}
  + \dfracp{}{\theta} \left\{ \left[ \omega + v[F] + H(\theta, t; \oex) \right] F \right\} = 0,
  \label{eq:EOC}
\end{equation}
where
\begin{equation}
  v[F](\theta,t;\tau) = \int_{-\infty}^{\infty} d\omega \int_{0}^{2\pi} d\theta
  ~ \Gamma(\theta-\theta') F(\theta',\omega,t-\tau).
\end{equation}
Suppose that the nonsynchronized state $F_{0}(\omega)=g(\omega)/(2\pi)$
is stable stationary under $H\equiv 0$. We expand $F$ around $F_{0}$ as
\begin{equation}
  F(\theta,\omega,t)
  = F_{0}(\omega) + f^{(1)}(\theta,\omega,t) + f^{(2)}(\theta,\omega,t) + \cdots,
  \label{eq:F-expansion}
\end{equation}
where $f^{(k)}=O(\norm{H}^{k})$.
Substituting the expansion \eqref{eq:F-expansion}
into the equation of continuity \eqref{eq:EOC}, we have
\begin{equation}
  \dfracp{f^{(1)}}{t} + \dfracp{}{\theta} \left[
    \omega f^{(1)} + \left( v[f^{(1)}] + H \right) F_{0} 
  \right] = 0
  \label{eq:EOC-H1}
\end{equation}
in the order of $O(\norm{H})$, and
\begin{equation}
  \dfracp{f^{(2)}}{t} + \dfracp{}{\theta} \left[
    \omega f^{(2)}+ v[f^{(2)}] F_{0} + \left( v[f^{(1)}] + H \right) f^{(1)} 
  \right] = 0
  \label{eq:EOC-H2}
\end{equation}
in the order of $O(\norm{H}^{2})$.
We analyze \eqref{eq:EOC-H1} and \eqref{eq:EOC-H2}
through the Fourier series expansion in $\theta$
and the Laplace transform in $t$.

\subsection{Fourier series expansion}

The coupling function $\Gamma$, the external force $H$,
and the perturbations $f^{(k)}$ are $2\pi$-periodic functions with respect to $\theta$,
and they are expanded into the Fourier series as
\begin{align}
  \Gamma\left(\theta\right)
  = -\sum_{m=1}^{\infty}K_{m}\sin\left(m\theta+\alpha_{m}\right)
  = - \sum_{n\neq 0} \Gamma_{n} e^{in\theta},
  \label{eq:coupling_functions}
\end{align}
\begin{equation}
  \begin{split}
    H\left(\theta, t; \oex \right)
    = - \Theta(t) \sum_{m=1}^{\infty} h_{m}\sin\left[m\left(\theta-\oex t\right)\right] 
    = - \sum_{n\neq 0} e^{in\theta} H_{n}(t;\oex),  \\
  \end{split}
  \label{eq:external_force}
\end{equation}
and
\begin{equation}
  f^{(k)}(\theta,\omega,t) = \sum_{n\neq 0} e^{in\theta} f^{(k)}_{n}(\omega,t).
\end{equation}
Here, we have the relations
\begin{equation}
  \Gamma_{n} = i \dfrac{K_{n}}{2}e^{i\alpha_{n}},
  \quad
  \Gamma_{-n}=\Gamma_{n}^{\ast}
  \quad
  (n>0)
\end{equation}
and
\begin{equation}
  H_{n}(t;\oex ) = i \dfrac{h_{n}}{2} \Theta(t) e^{-in\oex t},
  \quad
  H_{-n} = H_{n}^{\ast}
  \quad
  (n>0)
\end{equation}
where the superscript $\ast$ represents the complex conjugate.
We assume that $\Gamma_{0}=0$, since it is renormalized into $\omega$,
in other words, into a shift of the natural frequency distribution $g(\omega)$.
Note that there is no external force of the zeroth mode: $H_{0}\equiv 0$.
The order parameter functionals $z_{n}[f]$'s are defined by
\begin{equation}
  \begin{split}
    z_{n}[f](t)
    = \int_{-\infty}^{\infty} d\omega \int_{0}^{2\pi} d\theta~
    e^{in\theta} f(\theta,\omega,t)
    = 2\pi \int_{-\infty}^{\infty} f_{-n}(\omega,t).
  \end{split}
\end{equation}

The Fourier series expansions give
\begin{equation}
  \dfracp{f^{(1)}_{n}}{t} + in \left\{
    \omega  f^{(1)}_{n} + \left[ \Gamma_{n} z_{-n}^{(1)}(t-\tau) + H_{n} \right] F_{0} 
  \right\} = 0
\end{equation}
in $O(\norm{H})$ and
\begin{equation}
  \dfracp{f^{(2)}_{n}}{t} + in \left\{
    \omega  f^{(2)}_{n} + \Gamma_{n} z_{-n}^{(2)}(t-\tau) F_{0}
    + N_{n}^{(2)}
  \right\} = 0
\end{equation}
in $O(\norm{H}^{2})$. The symbol $z_{-n}^{(k)}(t)=z_{-n}[f^{(k)}](t)$ was introduced
to simplify the notation.
The second-order nonlinear term $N_{n}^{(2)}$ is defined by
\begin{equation}
  N_{n}^{(2)}(\omega,t)
  = \sum_{m} \left[ \Gamma_{m} z_{-m}^{(1)}(t-\tau) + H_{m}(t) \right] f^{(1)}_{n-m}(\omega,t).
\end{equation}

\subsection{Laplace transform}

From now on, the Laplace transform of a function is indicated by the upper hat symbol.
For an arbitrary analytic function $\varphi(t)$, the Laplace transform is defined by
\begin{equation}
  \wh{\varphi}(s) = \int_{0}^{\infty} e^{-st} \varphi(t) dt,
  \quad
  {\rm Re}(s)>0,
  \label{eq:Laplace-transform}
\end{equation}
where the domain ${\rm Re}(s)>0$ is introduced to ensure the convergence
of integral.
The perturbation $f$ is zero at $t=0$,
since $F_{0}$ is stable stationary and no external force is applied in $t<0$.
We hence have the Laplace transformed equations as
\begin{equation}
  (s+in\omega) \wh{f}^{(1)}_{n} + in \left(
    \Gamma_{n} e^{-s\tau} \wh{z}_{-n}^{(1)} + \wh{H}_{n}
  \right) F_{0} = 0
  \label{eq:Laplace-H1}
\end{equation}
in $O(\norm{H})$ and
\begin{equation}
  (s+in\omega) \wh{f}^{(2)}_{n} + in \left(
    \Gamma_{n} e^{-s\tau} \wh{z}_{-n}^{(2)} F_{0}
    + \wh{N}_{n}^{(2)}
  \right)
  = 0
  \label{eq:Laplace-H2}
\end{equation}
in $O(\norm{H}^{2})$.

\subsection{Linear response : $O(\norm{H})$}

The equation \eqref{eq:Laplace-H1} is solved algebraically.
Dividing $s+in\omega$, multiplying by $2\pi$, and integrating over $\omega$,
we have
\begin{equation}
  \wh{z}_{-n}^{(1)}(s) = - \dfrac{\wh{H}_{n}(s)}{\Lambda_{n}(s)} I_{n}(s),
  \quad
  {\rm Re}(s)>0.
  \label{eq:z-n1s}
\end{equation}
where the spectrum function $\Lambda_{n}(s)~(n\neq 0)$ is
\begin{equation}
  \Lambda_{n}(s) = 1 + \Gamma_{n} e^{-s\tau} I_{n}(s),
  \quad
  {\rm Re}(s)>0.
  \label{eq:Lambdan}
\end{equation}
and the integral $I_{n}(s)$ is
\begin{equation}
  I_{n}(s) = \int_{-\infty}^{\infty} \dfrac{g(\omega)}{\omega-is/n},
  \quad
  {\rm Re}(s)>0.
  \label{eq:In}
\end{equation}
The domain ${\rm Re}(s)>0$ comes from the domain of the Laplace transform
\eqref{eq:Laplace-transform}.

$I_{n}(s)$, and $\Lambda_{n}(s)$ and $z_{-n}^{(1)}(s)$ accordingly,
are analytically continued
to the whole complex $s$ plane as follows.
The integrand of $I_{n}(s)$ has the singularity at $\omega=is/n$,
which is located on the upper (lower) half of the complex $\omega$ plane
for ${\rm Re}(s)>0$ and $n>0$ ($n<0$).
Moving the singularity to the other half,
we smoothly modify the integral contour, the real axis,
so as to avoid the singularity.
As a result, the residue is added,
because the modified contour, denoted by ${\rm L}$,
encloses the singularity entirely for ${\rm Re}(s)<0$ and half for ${\rm Re}(s)=0$.
The continued integral $I_{n}(s)$ is therefore
\begin{equation}
  \begin{split}
    I_{n}(s)
    = \int_{\rm L} \dfrac{g(\omega)}{\omega-is/n} d\omega =
    \left\{
      \begin{array}{ll}
      \displaystyle{ \int_{-\infty}^{\infty} \dfrac{g(\omega)}{\omega-is/n} d\omega }
      & ({\rm Re}(s)>0) \\
      \displaystyle{ {\rm PV} \int_{-\infty}^{\infty} \dfrac{g(\omega)}{\omega-is/n} d\omega }
      + {\rm sgn}(n) i\pi g(is/n) 
      & ({\rm Re}(s)=0) \\
      \displaystyle{ \int_{-\infty}^{\infty} \dfrac{g(\omega)}{\omega-is/n} d\omega }
      + {\rm sgn}(n) i2\pi g(is/n) 
      & ({\rm Re}(s)<0) \\
    \end{array}
  \right.
  \end{split}
\end{equation}
where ${\rm PV}$ represents the Cauchy principal value,
and ${\rm sgn}(n)$ is the sign of $n$
representing the direction of the integral counter enclosing the singularity.

Temporal evolution of $z_{-n}^{(1)}(t)$ is obtained by performing
the inverse Laplace transform as
\begin{equation}
  z_{-n}^{(1)}(t) = \dfrac{1}{2\pi i} \int_{\sigma-i\infty}^{\sigma+i\infty}
  e^{st} \wh{z}_{-n}^{(1)}(s) ds,
\end{equation}
where $\sigma\in\mathbb{R}$ is larger than the real parts of any singularities of
$\wh{z}_{-n}^{(1)}(s)$.
The continuation of $\wh{z}_{-n}^{(1)}(s)$ permits us to use the residue theorem
by adding the half-circle lying in left-half of the complex $s$ plane;
The inverse Laplace transform picks up the singularity of $\wh{z}_{-n}^{(1)}(s)$.
The asymptotic behavior is determined by the pole of $\wh{z}_{-n}(s)$
which has the largest real part.
Since we assumed that the reference state $F_{0}$ is stable,
all the roots of $\Lambda_{n}(s)$ are  in the region ${\rm Re}(s)<0$,
which induce the Landau damping.
The asymptotic behavior is hence determined by the poles of
$\wh{H}_{n}(s)$ and $\wh{H}_{-n}(s)$, which are
\begin{equation}
  \wh{H}_{n}(s) = \dfrac{ih_{n}}{2} \dfrac{1}{s+in\oex},
  \quad
  \wh{H}_{-n}(s) = \dfrac{-ih_{n}}{2} \dfrac{1}{s-in\oex},
  \quad (n>0).
  \label{eq:Hns}
\end{equation}
The continued integrals $I_{n}(s)$ at the poles are 
\begin{equation}
  I_{n}(-in\oex)
  = i \mathcal{G}^{\ast}(\oex),
  \quad
  I_{-n}(in\oex)
  = -i \mathcal{G}(\oex),
  \quad
  (n>0)
\end{equation}
where
\begin{equation}
  \mathcal{G}(\oex)
  = \pi g(\oex) + i {\rm PV} \int_{-\infty}^{\infty}
  \dfrac{g(\omega)}{\omega-\oex} d\omega.
\end{equation}
The spectrum functions at the poles are 
\begin{equation}
  \Lambda_{n}(-in\oex) = \dfrac{1}{2} \left[ 2 - L_{n}^{\ast} \mathcal{G}^{\ast}(\oex) \right],
  \quad
  \Lambda_{-n}(in\oex) = \dfrac{1}{2} \left[ 2 - L_{n} \mathcal{G}(\oex) \right],
  \quad
  (n>0)
  \label{eq:Lambdan-inoex}
\end{equation}
where $L_{n}=K_{n}e^{-i(\alpha_{n}+n\oex\tau)}$.

Putting all together, the asymptotic temporal evolution is for $n>0$ is
\begin{equation}
  z_{-n}^{(1)}(t) \xrightarrow{t\to\infty}
  e^{-in\oex t} \dfrac{\mathcal{G}^{\ast}(\oex)}{2-L_{n}^{\ast}\mathcal{G}^{\ast}(\oex)} h_{n},
  \quad
  z_{n}^{(1)}(t) \xrightarrow{t\to\infty}
  e^{in\oex t} \dfrac{\mathcal{G}(\oex)}{2-L_{n}\mathcal{G}(\oex)} h_{n}.
\end{equation}
The susceptibility $\chi_{n}^{m}(\oex)$ defined by
\begin{equation}
  e^{-in\oex t} z_{n}^{(1)}(t) \xrightarrow{t\to\infty}
  \sum_{m} \chi_{n}^{m}(\oex) h_{m} + O(\norm{H}^{2}),
  \quad
  e^{in\oex t} z_{-n}^{(1)}(t) \xrightarrow{t\to\infty}
  \sum_{m} \chi_{-n}^{-m}(\oex) h_{-m} + O(\norm{H}^{2}),
  \quad
  (n>0)
\end{equation}
is hence
\begin{equation}
  \chi_{n}^{m}(\oex) = \chi_{n}(\oex) \delta_{nm},
  \quad
  \chi_{-n}^{-m}(\oex) = \chi_{-n}(\oex) \delta_{nm}, 
  \quad
  (n>0),
\end{equation}
where 
\begin{equation}
  \chi_{n}(\oex) = \dfrac{\mathcal{G}(\oex)}{2-L_{n}\mathcal{G}(\oex)},
  \quad
  \chi_{-n}(\oex) = \dfrac{\mathcal{G}^{\ast}(\oex)}{2-L_{n}^{\ast}\mathcal{G}^{\ast}(\oex)},
  \quad
  (n>0).
  \label{eq:chin}
\end{equation}

\subsection{Nonlinear response : $O(\norm{H}^{2})$}
The same way as $O(\norm{H})$ gives the Laplace transform $\wh{z}_{-n}^{(2)}(s)$ as
\begin{equation}
  \wh{z}_{-n}^{(2)}(s)
  = \dfrac{-2\pi}{\Lambda_{n}(s)}
  \int_{-\infty}^{\infty} \dfrac{\wh{N}_{n}^{(2)}(\omega,s)}{\omega-is/n} d\omega.
\end{equation}
We need the Laplace transform of products, which appear in $\wh{N}_{n}^{(2)}$.

\subsubsection{Laplace transform of a product function}
\label{sec:laplace-product}
For analytic functions $f(t)$ and $g(t)$, we have the relation
\begin{equation} 
  \wh{fg}(s) = \dfrac{1}{2\pi i} \int_{\sigma_{g}-i\infty}^{\sigma_{g}+i\infty} \wh{f}(s-s') \wh{g}(s') ds',
  \label{eq:convolution}
\end{equation}
where $\sigma_{g}\in\mathbb{R}$
is larger than the real parts of any singularities of $\wh{g}(s)$.
A proof of \eqref{eq:convolution} is straightforward.
We denote the inverse Laplace transforms of $\wh{f}(s)$ and $\wh{g}(s)$ as
\begin{equation}
  f(t)
  = \dfrac{1}{2\pi i} \int_{\sigma_{f}-i\infty}^{\sigma_{f}+i\infty} e^{s_{1}t} \wh{f}(s_{1}) ds_{1},
\end{equation}
where $\sigma_{f}\in\mathbb{R}$
is larger than the real parts of any singularities of $\wt{f}(s)$, and
\begin{equation}
  g(t)
  = \dfrac{1}{2\pi i} \int_{\sigma_{g}-i\infty}^{\sigma_{g}+i\infty} e^{s_{2}t} \wh{g}(s_{2}) ds_{2}.
\end{equation}
Changing the variables as $(s,s')=(s_{1}+s_{2},s_{2})$,
the product function $(fg)(t)$ is expressed as
\begin{equation}
  \begin{split}
    (fg)(t) =
    \dfrac{1}{2\pi i}
    \int_{\sigma_{f}+\sigma_{g}-i\infty}^{\sigma_{f}+\sigma_{g}+i\infty} ds~ e^{st} 
    \left[ \dfrac{1}{2\pi i} \int_{\sigma_{g}-i\infty}^{\sigma_{g}+i\infty} ds'
     \wh{f}(s-s') \wh{g}(s') \right].
  \end{split}
\end{equation}
The integral over $s$ is the inverse Laplace transform of the inside of
the square brackets,
and hence we have the relation \eqref{eq:convolution}.

We note that we pick up the singularities of $\wh{g}$ only
in the integral with respect to $s'$. 
Let $a$ be a pole of $\wh{f}(s)$, and $b$ of $\wh{g}(s)$.
By the definitions, we have ${\rm Re}(a)<\sigma_{1}$ and ${\rm Re}(b)<\sigma_{2}$.
The convolution yields a pole of $\wh{f}$ which lies on the right-side of the line
${\rm Re}(s')=\sigma_{g}$, since $s'=s-a=\sigma_{f}+\sigma_{g}-a>\sigma_{g}$.
Therefore, this singularity is not enclosed by the integral counter,
which consists of the line ${\rm Re}(s')=\sigma_{g}$
and the left half-circle passing through the point at infinity
on the left-half complex $s'$ plane.

\subsubsection{Convolution in $\wh{N}_{n}^{(2)}$}

Let us denote
\begin{equation}
  V_{m}(t) = \Gamma_{m}z_{-m}^{(1)}(t-\tau)+H_{m}(t),
\end{equation}
which rewrite the nonlinear term $N_{n}^{(2)}$ into
\begin{equation}
  N_{n}^{(2)}(\omega,t) = \sum_{m} V_{m}(t) f_{n-m}^{(1)}(\omega,t).
\end{equation}
The Laplace transform $\wh{z}_{-n}^{(2)}(s)$ is expressed as
\begin{equation}
  \wh{z}_{-n}^{(2)}(s)
  = \dfrac{-2\pi}{\Lambda_{n}(s)}
  \sum_{m} \int_{-\infty}^{\infty} \dfrac{\mathcal{L}[V_{m}f_{n-m}^{(1)}](s)}{\omega-is/n} d\omega,
  \label{eq:z-n2s}
\end{equation}
where $\mathcal{L}$ represents the Laplace transform operator.

The Laplace transform of $V_{m}$ is
\begin{equation}
  \wh{V}_{m}(s)
  = \Gamma_{m} e^{-s\tau} \wh{z}_{-m}^{(1)}(s) + \wh{H}_{m}(s)
  = \dfrac{\wh{H}_{m}(s)}{\Lambda_{m}(s)},
\end{equation}
where we used \eqref{eq:z-n1s} and \eqref{eq:Lambdan}.
The Laplace transform $\wh{f}_{m}^{(1)}(\omega,s)$ is then
from \eqref{eq:Laplace-H1}
\begin{equation}
  \wh{f}_{m}^{(1)}(\omega,s)
  = - \dfrac{F_{0}(\omega)}{\omega-is/m} \dfrac{\wh{H}_{m}(s)}{\Lambda_{m}(s)}.
\end{equation}

The Laplace transform of $V_{m}f_{n-m}^{(1)}$ is
\begin{equation}
  \begin{split}
    \mathcal{L}[V_{m}f_{n-m}^{(1)}](s) 
    = \dfrac{1}{2\pi i} \int_{\sigma_{2}-i\infty}^{\sigma_{2}+i\infty}
    \dfrac{\wh{H}_{m}(s')}{\Lambda_{m}(s')}
    \dfrac{F_{0}(\omega)}{\omega-i\frac{s-s'}{n-m}}
    \dfrac{\wh{H}_{n-m}(s-s')}{\Lambda_{n-m}(s-s')} ds'.
  \end{split}
\end{equation}
Remembering the note at the end of Sec.~\ref{sec:laplace-product}
and keeping in mind that we are interested in the asymptotic temporal evolution,
we pick up the pole of $\wh{H}_{m}(s')$ which is at $s'=-im\oex$.
The principal part of the Laplace transform is then
\begin{equation}
  \begin{split}
    {\rm PP} \mathcal{L}[V_{m}f_{n-m}^{(1)}](s) 
    = \dfrac{{\rm Res}(\wh{H}_{m})}{\Lambda_{m}(-im\oex)}
    \dfrac{\wt{H}_{n-m}(s+im\oex)}{\Lambda_{n-m}(s+im\oex)}
    \dfrac{F_{0}(\omega)}{\omega-i\frac{s+im\oex}{n-m}},
  \end{split}
\end{equation}
where PP represents the principal part surviving in the limit $t\to\infty$,
and ${\rm Res}(\wh{H}_{m})={\rm sgn}(m)ih_{m}/2$ is the residue of $\wh{H}_{m}$.
Substituting the above expression into \eqref{eq:z-n2s}, we have
\begin{equation}
  \begin{split}
    {\rm PP} \wh{z}_{-n}^{(2)}(s)
    = \dfrac{-1}{\Lambda_{n}(s)} \sum_{m}
    \dfrac{{\rm Res}(\wh{H}_{m})}{\Lambda_{m}(-im\oex)}
    \dfrac{\wh{H}_{n-m}(s+im\oex)}{\Lambda_{n-m}(s+im\oex)} T_{n,m}(s),
    \label{eq:z-n2s}
  \end{split}
\end{equation}
where
\begin{equation}
  T_{n,m}(s)
  = \int_{\rm L} \dfrac{g(\omega)}
  {\left( \omega-i\frac{s+im\oex}{n-m} \right)\left( \omega-i\frac{s}{n}\right)} d\omega.
\end{equation}
We pick up the pole of $\wh{H}_{n-m}(s+im\oex)$, which is at $s=-in\oex$,
for the asymptotic temporal evolution. Then,
\begin{equation}
  \begin{split}
    e^{in\oex t} z_{-n}^{(2)}(t) \xrightarrow{t\to\infty}
    \dfrac{-1}{\Lambda_{n}(-in\oex)}
    \sum_{m}
    \dfrac{{\rm Res}(\wh{H}_{m}){\rm Res}(\wh{H}_{n-m})T_{n,m}(-in\oex)}
    {\Lambda_{m}(-im\oex)\Lambda_{n-m}(-i(n-m)\oex)}.
    \label{eq:z-n2t}
  \end{split}
\end{equation}
We have to be careful for the value $T_{n,m}(-in\oex)$,
because the integrand of $T_{n,m}(-in\oex)$
has the pole of order two at $\omega=\oex$.

\subsubsection{Nonlinear response coefficient}

From now on, we focus on the linear response of the mode $2$
induced by the external force of the mode $1$, i.e. $h_{1}>0$ and $h_{l}=0~(l>1)$.
Setting $n=2$ and $m=1$ in \eqref{eq:z-n2t}, we have
\begin{equation}
  e^{2i\oex t} z_{-2}^{(2)}(t) \xrightarrow{t\to\infty}
  \dfrac{T_{2,1}(-2i\oex)}
  {4\Lambda_{2}(-2i\oex)[\Lambda_{1}(-i\oex)]^{2}} h_{1}^{2}.
\end{equation}
To obtain the value $T_{2,1}(-2i\oex)$,
we first perform the partial fraction decomposition as
\begin{equation}
  T_{2,1}(s)
  = \dfrac{2}{i(s+2i\oex)}
  \left[  I_{1}(s+i\oex) - I_{2}(s) \right].
\end{equation}
In the limit $s\to -2i\oex'~(\oex'\neq\oex)$ from the upper-half $s$ plane, we have
\begin{equation}
  T_{2,1}(-2i\oex') = \dfrac{i}{\oex'-\oex}
  \left[ \mathcal{G}^{\ast}(2\oex'-\oex) - \mathcal{G}^{\ast}(\oex') \right].
\end{equation}
Further taking the limit $\oex'\to \oex$, we have
\begin{equation}
  T_{2,1}(-2i\oex) = i \left( \mathcal{G}^{\ast} \right)' (\oex).
\end{equation}
The asymptotic temporal evolution of $z_{2}^{(2)}(t)$ is hence
\begin{equation}
  e^{-2i\oex t} z_{2}^{(2)}(t) \xrightarrow{t\to\infty}
  \chi_{2}^{11}(\oex) h_{1}^{2} + O(\norm{H}^{3}),
\end{equation}
where
\begin{equation}
  \chi_{2}^{11}(\oex) = 
  \dfrac{i \mathcal{G}'(\oex)}
  {4\Lambda_{2}^{\ast}(-2i\oex)[\Lambda_{1}^{\ast}(-i\oex)]^{2}}.
\end{equation}
Substituting \eqref{eq:Lambdan-inoex} into the above expression,
we have
\begin{equation}
  \begin{split}
    \chi_{2}^{11}(\oex)
    = \dfrac{2i \mathcal{G}'(\oex)}
    {[2-L_{2}(\oex)\mathcal{G}(\oex)][2-L_{1}(\oex)\mathcal{G}(\oex)]^{2}} 
    = \dfrac{2i \mathcal{G}'(\oex)}{[\mathcal{G}(\oex)]^{3}}
    \chi_{2}(\oex) [\chi_{1}(\oex)]^{2},
    \\
  \end{split}
  \label{eq:chi211}
\end{equation}
where we used \eqref{eq:chin}.

\section{Inference of $L_{1}$}

The nonlinear response coefficient \eqref{eq:chi211} gives
\begin{equation}
  \mathcal{G}'(\oex)
   = \dfrac{\chi_{2}^{11}(\oex)[\mathcal{G}(\oex)]^{3}}{2i\chi_{2}(\oex)[\chi_{1}(\oex)]^{2}}.
  \label{eq:Gdash-1}
\end{equation}
Another expression of $\mathcal{G}'(\oex)$ is obtained by solving
\eqref{eq:chin} by $\mathcal{G}(\oex)$ as
\begin{equation}
  \mathcal{G}(\oex) = \dfrac{2\chi_{n}(\oex)}{1+L_{n}(\oex)\chi_{n}(\oex)}
  \label{eq:mathcalG}
\end{equation}
and derivating it with respect to $\oex$ as
\begin{equation}
  \begin{split}
    \mathcal{G}'(\oex)
    = 2 \dfrac{\chi_{n}'[1+L_{n}\chi_{n}]-\chi_{n}[L_{n}\chi_{n}]'}{[1+L_{n}\chi_{n}]^{2}} 
    = \dfrac{\chi_{n}'(\oex) +in\tau L_{n}[\chi_{n}(\oex)]^{2}}{2[\chi_{n}(\oex)]^{2}} [\mathcal{G}(\oex)]^{2}. \\
  \label{eq:Gdash-2}
  \end{split}
\end{equation}
where we used the definition $L_{n}=K_{n}e^{-i(\alpha_{n}+n\oex\tau)}$.
The combination between \eqref{eq:Gdash-1} and \eqref{eq:Gdash-2}
provides for $n=1$
\begin{equation}
  \mathcal{G}(\oex)
  = \dfrac{i\chi_{2}(\oex)[\chi_{1}'(\oex) +i\tau L_{1}[\chi_{1}(\oex)]^{2}]}
  {\chi_{2}^{11}(\oex)}.
\end{equation}
This expression and \eqref{eq:mathcalG} for $n=1$ give the equality
\begin{equation}
  \dfrac{1+L_{1}(\oex)\chi_{1}(\oex)}{2\chi_{1}(\oex)}
  = \dfrac{\chi_{2}^{11}(\oex)}
  {i\chi_{2}(\oex)\{\chi_{1}'(\oex)+i\tau L_{1}[\chi_{1}(\oex)]^{2}\}}.
  \label{eq:L1-quadratic}
\end{equation}
This is the equation for determining $L_{1}$.

\subsection{For $\tau=0$}

In particular, $L_{1}$ is uniquely determined for $\tau=0$ as
\begin{equation}
  L_{1}
  = K_{1}e^{-i\alpha_{1}}
  = \dfrac{2\chi_{2}^{11}(\oex)}{i\chi_{2}(\oex)\chi_{1}'(\oex)}  - \dfrac{1}{\chi_{1}(\oex)}.
  \label{eq:L1-tau0}
\end{equation}

\subsection{For $\tau>0$}
We can infer $L_{1}$ from the quadratic equation \eqref{eq:L1-quadratic}
for $\tau>0$ as well as for $\tau=0$.
The quadratic equation is rewritten into
\begin{equation}
  A L_{1}^{2} + B L_{1} + C = 0,
  \label{eq:L1-quadratic-ABC}
\end{equation}
where
\begin{equation}
  A(\oex) = i \tau \dfrac{[\chi_{1}(\oex)]^{2}}{\chi_{1}'(\oex)},
  \quad
  B(\oex) = 1 + i \tau \dfrac{\chi_{1}(\oex)}{\chi_{1}'(\oex)},
  \quad
  C(\oex) = \dfrac{1}{\chi_{1}(\oex)} - \dfrac{2\chi_{2}^{11}(\oex)}{i\chi_{2}(\oex)\chi_{1}'(\oex)}.
\end{equation}
We have the two solutions to \eqref{eq:L1-quadratic-ABC},
and we select the solution
\begin{equation}
  L_{1}(\oex) = - \dfrac{B(\oex)}{2A(\oex)} \left( 1 - \sqrt{1 - \dfrac{4A(\oex)C(\oex)}{[B(\oex)]^{2}}} \right)
  \label{eq:L1-Method-2}
\end{equation}
to have \eqref{eq:L1-tau0} in the limit $\tau\to 0$, namely $A\to 0$.
The inferred $L_{1}$ induces the other inferences of $L_{m}$'s through the relation
\begin{equation}
  L_{m}(\oex) - L_{1}(\oex) = \dfrac{1}{\chi_{1}(\oex)} - \dfrac{1}{\chi_{m}(\oex)}
  \quad
  (m\geq 2).
  \label{eq:Lm-relation}
\end{equation}

The inferred parameter values are summarized in Table \ref{tab:inference}
for {\bf Model-1}.
The inferred coupling function $\Gamma_{1}(\theta)$
and the natural frequency distribution $g_{1}(\omega)$
are compared with the true ones in Fig.~\ref{fig:Example2-Method2}.
We observe rather large errors in higher order modes in $\Gamma_{1}(\theta)$,
and precision is improved by truncating the Fourier series up to the mode-$3$.
Moreover, the errors tend to decrease as the number of samples increases,
and $g_{1}(\omega)$ is well inferred irrespective of used modes.

\begin{table}[h]
  \centering
  \caption{True and inferred parameter values of {\bf Model-1}
    from \eqref{eq:L1-Method-2} and \eqref{eq:Lm-relation},
    by taking the average over $\oex$.
    The time delay $\tau$ is inferred by {\bf Procedure-1}.  }
  \begin{tabular}{l|l|llllllllll}
    \hline
    {\bf Model-1} & $\tau$ & $K_{1}$ & $\alpha_{1}$ & $K_{2}$ & $\alpha_{2}$ & $K_{3}$ & $\alpha_{3}$ & $K_{4}$ & $\alpha_{4}$ & $K_{5}$ & $\alpha_{5}$ \\
    \hline
    Truth & 2 & 1.379 & 0.7884 & 0.568 & -3.0316 & 0.154 & -0.7546 & 0 & -- & 0 & -- \\
    $\Omega_{1}^{50}$ & 1.987 & 1.215 & 0.925 & 0.683 & -2.663 & 0.257 & 0.694 & 0.119 & 2.108 & 0.289 & 0.991 \\
    $\Omega_{1}^{25}$ & 1.995 & 0.857 & 0.806 & 0.956 & -2.584 & 0.414 & 1.004 & 0.253 & 1.190 & 0.389 & 0.407 \\
    \hline
  \end{tabular}
  \label{tab:inference}
\end{table}

\begin{figure}[h]
  \centering
  \includegraphics[width=16cm]{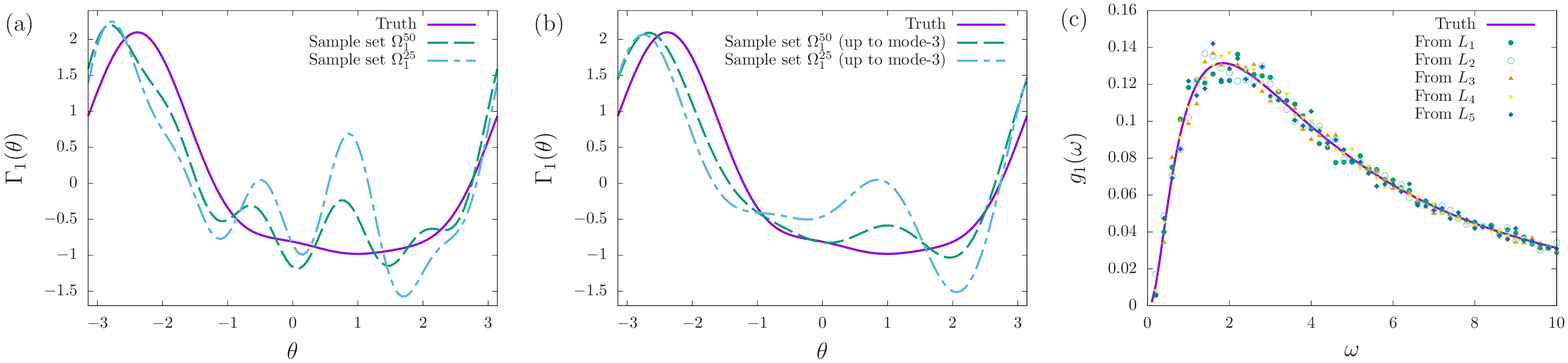}
  \caption{Comparison between the truth and the inference
    in {\bf Model-1} having $\tau>0$.
    (a) The coupling function $\Gamma_{1}(\theta)$
    produced from the sample set $\Omega_{1}^{50}$ (green broken line),
    $\Omega_{1}^{25}$ (blue chain line).
    (b) Same as (a) but the inferred $\Gamma_{1}(\theta)$ are truncated
    up to the Fourier mode-$3$.
    (c) The natural frequency distribution $g_{1}(\omega)$
    obtained from the inferred $L_{1}$ (green filled circles),
    $L_{2}$ (blue open circles),
    $L_{3}$ (orange triangles),
    $L_{4}$ (yellow inverse triangles),
    and $L_{5}$ (dark-blue diamonds).
    The sample set is $\Omega_{1}^{50}$.}
  \label{fig:Example2-Method2}
\end{figure}

\end{widetext}


\begin{thebibliography}{99}

\bibitem{winfree-01}
  A. T. Winfree,
  {\it The Geometry of Biological Time}
  (Springer, New York, 2001).
  
\bibitem{strogatz-03}
  S. H. Strogatz, 
  {\it Sync: How order emerges from chaos in the universe, nature, and daily life}
  (Hyperion, New York, 2003).
 
\bibitem{pikovsky-01}
  A. Pikovsky, M. Rosenblum, and J. Kurths,
  {\it Synchronization: a universal concept in nonlinear sciences}
  (Cambridge University Press, Cambridge, 2001).
  
 \bibitem{palmigiano-17} 
A. Palmigiano, T. Geisel, F. Wolf, and D. Battaglia,
 Flexible information routing by transient synchrony,
  Nat. Neurosci. {\bf 20}, 1014-1022 (2017).

\bibitem{buzsaki-13}
G. Buzs{\'a}ki and E.I. Moser,
 Memory, navigation and theta rhythm in the hippocampal-entorhinal system,
  Nat. Neurosci. {\bf 16}, 130-138 (2013).
  
\bibitem{kuramoto-03}
  Y. Kuramoto,
  {\it Chemical oscillations, waves, and turbulence}
  (Dover, New York, 2003).

\bibitem{nakao-16}
   H. Nakao,
   Phase reduction approach to synchronisation of nonlinear oscillators,
   Contemp. Phys. {\bf 57}, 188 (2016).

\bibitem{kuramoto-19}
  Y. Kuramoto and H. Nakao,
  On the concept of dynamical reduction: the case of coupled oscillators,
  Phil. Trans. R. Soc. A {\bf 377}, 20190041 (2019).
  

\bibitem{galan-ermentrout-urban-05}
  R. F. Gal{\'a}n, G. B. Ermentrout, and N. N. Urban,
  Efficient estimation of phase-resetting curves in real neurons and its significance for neural-network modeling,
  Phys. Rev. Lett. {\bf 94}, 158101 (2005).
  
\bibitem{miyazaki-kinoshita-05}
  J. Miyazaki and S. Kinoshita,
  Determination of a coupling function in multicoupled oscillators,
  Phys. Rev. Lett. {\bf 96}, 194101 (2005).

 \bibitem{tokuda-07}
   I. T. Tokuda, S. Jain, I. Z. Kiss, and J. L. Hudson,
   Inferring phase equations from multivariate time series,
   Phys. Rev. Lett. {\bf 99}, 064101 (2007).
   
 \bibitem{kralemann-07}
   B. Kralemann, L. Cimponeriu, M. Rosenblum, A. Pikovsky, and R. Mrowka,
   Uncovering interaction of coupled oscillators from data,
   Phys. Rev. E {\bf 76}, 055201(R) (2007).
   
 \bibitem{kralemann-08}
   B. Kralemann, L. Cimponeriu, M. Rosenblum, A. Pikovsky, and R. Mrowka,
   Phase dynamics of coupled oscillators reconstructed from data,
   Phys. Rev. E {\bf 77}, 066205 (2008).
   
 \bibitem{penny-09}
   W. D. Penny, V. Litvak, L. Fuentemilla, E. Duzel, and K. Friston,
   Dynamic Causal Models for phase coupling,
   J. Neurosci. Methods {\bf 183}, 19 (2009).
   
 \bibitem{stankovski-12}
   T. Stankovski, A. Duggento. P. V. E. McClintock, and A Stefanovska,
   Inference of Time-Evolving Coupled Dynamical Systems in the Presence of Noise,
   Phys. Rev. Lett. {\bf 109}, 024101 (2012).
   
 \bibitem{ota-14}
   K. Ota and T. Aoyagi,
   Direct extraction of phase dynamics from fluctuating rhythmic data based on a Bayesian approach,
   arXiv: 1405.4126 (2014).
   
 \bibitem{pikovsky-18}
   A. Pikovsky,
   Reconstruction of a random phase dynamics network from observations,
   Phys. Lett. A 382, 147 (2018).
   
\bibitem{mori-22}
   F. Mori and H. Kori,
   Noninvasive inference methods for interaction and noise intensities of coupled oscillators using only spike time data,
   Proc. Natl. Acad. Sci. {\bf 119}, e2113620119 (2022).

 \bibitem{yeung-strogatz-99}
   M. K. S. Yeung and S. H. Strogatz,
   Time delay in the Kuramoto model of coupled oscillators,
   Phys. Rev. Lett. {\bf 82}, 648 (1999).

 \bibitem{montbrio-pazo-schmidt-06}
   E. Montbri{\'o}, D. Paz{\'o}, and J. Schmidt,
   Time delay in the Kuramoto model with bimodal frequency distribution,
   Phys. Rev. E {\bf 74}, 056201 (2006).
   
    
\bibitem{sakaguchi-88}
   H. Sakaguchi,
   Cooperative Phenomena in Coupled Oscillator Systems under External Fields,
   Prog. Theor. Phys. {\bf 79}, 39 (1988).
  
 \bibitem{daido-15}
   H. Daido,
   Susceptibility of large populations of coupled oscillators,
   Phys. Rev. E {\bf 91} 012925 (2015).
   
 \bibitem{terada-yamaguchi-19}
   Y. Terada and Y. Y. Yamaguchi,
  Linear response theory for coupled phase oscillators with general coupling functions,
  J. Phys. A: Math. Theor. {\bf 53}, 044001 (2020).

\bibitem{watts-strogatz-98}
  D. J. Watts and S. H. Strogatz,
  Collective dynamics of `small-world' networks,
  Nature {\bf 393}, 440 (1998).

\bibitem{hong-choi-kim-02}
  H. Hong, M. Y. Choi, and B. J. Kim,
  Synchronization on small-world networks,
  Phys. Rev. E {\bf 65}, 026139 (2002).

\bibitem{yoneda-harada-yamaguchi-20}
  R. Yoneda, K. Harada, and Y. Y. Yamaguchi,
  Critical exponents in coupled phase-oscillator models on small-world networks,
  Phys. Rev. E {\bf 102}, 062212 (2020).
  
  \bibitem{daido-87}
  H. Daido,
  Population Dynamics of Randomly Interacting Self-Oscillators. I: Tractable Models without Frustration,
  Prog. Theor. Phys. {\bf 77}, 622 (1987).
  
  \bibitem{ichinomiya-03}
  T. Ichinomiya,
  Frequency synchronization in a random oscillator network,
  Phys. Rev. E {\bf 70}, 026116 (2004).
  
  \bibitem{hoppensteadt-97}
  F. C. Hoppensteadt and E.M. Izhikevich,
  {\it Weakly connected neural networks}
  (Springer, New York, 1997).

\bibitem{lancellotti-05}
  C. Lancellotti,
  On the Vlasov limit for systems of nonlinearly coupled oscillators without noise,
  Transport Theory and Statistical Physics {\bf 34}, 523 (2005).
 
\bibitem{daido-92}
  H. Daido,
  Order function and macroscopic mutual entrainment in uniformly coupled limit-cycle oscillators,
  Prog. Theor. Phys. {\bf 88}, 1213 (1992).

\bibitem{supplement}
  See the Supplementary Material at [URL].

\bibitem{hansel-93}
  D. Hansel, G. Mato, and C. Meunier,
  Phase Dynamics for Weakly Coupled Hodgkin-Huxley Neurons,
  Europhys. Lett. {\bf 23}, 367 (1993).
 
\bibitem{hansel-95}
  D. Hansel, G. Mato, and C. Meunier,
  Synchrony in excitatory neural networks,
  Neural Comput. {\bf 7}, 307-337 (1995).
 
  \bibitem{izhikevich-06}
  E. M. Izhikevich, 
  Polychronization: computation with spikes,
  Neural Comput. {\bf 18}  245 (2006).
  
\bibitem{buzsaki-14}
  G. Buzs{\'a}ki and K. Mizuseki,
  The log-dynamic brain: how skewed distributions affect network operations,
  Nat. Rev. Neurosci. {\bf 15}, 264 (2014).
  
  \bibitem{sakaguchi-86}
  H. Sakaguchi and Y. Kuramoto,
  A soluble active rotater model showing phase transitions via mutual entertainment,
  Prog. Theor. Phys.. {\bf 76}, 576 (1986).

  
\end{thebibliography}
\end{document}